# Protein Structure Prediction until CASP15

Arne Elofsson, Science for Life Laboratory and Dep. of Biochemistry and Biophysics, Stockholm University

## Abstract

In Dec 2020, the results of AlphaFold2 were presented at CASP14, sparking a revolution in the field of protein structure predictions. For the first time, a purely computational method could challenge experimental accuracy for structure prediction of single protein domains. The code of AlphaFold2 was released in the summer of 2021, and since then, it has been shown that it can be used to accurately predict the structure of most (ordered) proteins and many protein-protein interactions. It has also sparked an explosion of development in the field, improving AI-based methods to predict protein complexes, disordered regions, and protein design. Here I will review some of the inventions sparked by the release of AlphaFold.

## Background

Protein structure prediction is often divided into template-based and de-novo protein structure prediction. In the 1980s [1], methods were developed to predict the structure of a protein by copying the coordinates from an experimentally determined protein structure. Since the first methods, there have been methodological advances, but more importantly, the increase in sequence and structure data has enabled many more proteins to be modelled accurately [2].

Traditionally de novo modelling meant modelling a protein without using a template. However, this strict definition has blurred over the last two decades by using fragments. Methods such as Rosetta [3], I-Tasser [4], and FragFold [5] use fragments of various sizes to predict the structure of a protein. These fragments might come from homologous protein structures or not. Most successful in this area is the Rosetta program developed by the Baker group. However, only a small subset of proteins without templates could be modelled with any accuracy using this methodology, i.e. it is not very useful for general protein structure predictions.

In the 1990s, it was proposed that it could predict the structure of a protein by using co-evolutionary signals in a multiple sequence alignment [6]. However, the performance of these methods was minimal, and predictions were only slightly better than random. In 1999 a solution to increasing the accuracy of contact predictions was proposed by separating direct and indirect contacts using a trainable model [7]. A similar idea had been proposed earlier [8].  Unfortunately, these papers were largely ignored by the community. The methods could have been more helpful if only more protein families had sufficient members.

About ten years ago, the idea of indirect correlation was rediscovered [9], and other methods were also proposed [10].  These methods provided the ability to predict the structure of many proteins accurately [11]. However, the method was still limited to large protein families, and the models' accuracy was limited.

It was soon realised that combining DCA and machine (deep) learning was a way to improve the prediction of contacts [12]. The most successful method uses a very deep network to predict contacts with significantly better accuracy than DCA-based methods [13]. The requirements for very large multiple-sequence alignments also decreased.

At CASP13, DeepMind introduced AlphaFold (version 1) [14], building on the earlier methods. The main differences were that the architecture was deeper and that the network predicted distance probabilities, and not only contacts were predicted. This resulted in such accuracy that it was possible to use a simple steepest descent methodology to predict the structure of proteins. However, only a tiny fraction of proteins without templates could be predicted at experimental accuracies. The academic community reproduced the performance of AlphaFold1 quickly [15].

## AlphaFold v2.0

DeepMind presented AlphaFold2 at the CASP14. The results of AlphaFold2 were impressive. The average GDTts was close to 90 for individual domains, i.e. it was, in principle, as good as experimental structures. In the summer of 2021, the AlphaFold2 papers were published. In the first paper [16], the method was described and accompanied by a paper describing an extensive database with predicted structures [17]. This database has recently been extended to cover virtually all known proteins.

Notably, DeepMind released the source code of AlphaFold with an open-source license. This allowed the community to test and extend the method, spurring a real jump in development speed. In short, AlphaFold2 consists of two main modules the EvoFormer block and the structure block. Both of these blocks contain important innovations inspired by previous work. Most importantly, AlphaFold2 can be seen as an awe-inspiring engineering feat.

In contrast to AlphaFold1, the input to AlphaFold2 is a “raw” multiple sequence alignment, i.e. the deep learning network extracts co-evolutionary (and other) types of information directly from the MSA as proposed earlier [18].

A representation of the rawMSA (and a pair representation) is then used to input the 48 evoformer blocks. The Evoformer block is a two-track network, where first, a row and column-wise attention mechanism is used to analyse the MSA. This is then transformed into a pairwise description by an outer product, which is then updated using self-attention. One key innovation here is the triangle updates that can learn not to break triangle equalities. The output of one EvoFormer module is fed into the next. The final pair representation is used as an input to the structural module jointly with the representative sequence.

Notably, the structure module in AlphaFold2 is locally translated and rotational equivariant [19]. It starts treating the protein as a residue gas, i.e. ignoring all the chemistry from a protein being a polymer. This choice makes it ideal for treating proteins with missing residues and was one of the fundamental reasons why it was so easy to transform AlphaFold2 to work on multimers. Each residue is modelled as a triangle, and their internal relationships are described using affine matrices. One key invention here is the Invariant Point Attention (IPA), making the structure module invariant to translation and rotations (which is required as rotations and translations are arbitrary for a single protein). It is more straightforward (and therefore easier to learn) than e.g. SE(3) transformation. IPA is based on the L2-norm of a vector to be invariant concerning translations and rotations. The structure module (in contrast to other methods) also contains information about sidechains, represented by standard dihedral angles. Also, using Frame Aligned Point Error (instead of RMSD) as a loss function is novel and ingenious. Earlier attempts to develop an end-to-end method were unsuccessful because the structural representation was not optimal [20]. Even methods implemented after AlphaFold use less efficient methods [21,22] and often only predict backbone coordinates and, therefore, require external programs to generate all-atom models

# The field after the release of AlphaFold2

Thanks to a detailed presentation of the algorithm at the CASP conference and, more importantly, the release of the code as open source in June 2021, AlphaFold was the start of tremendous activity in the entire community (which also expanded rapidly). Many labs use the AlphaFold2 paper and the code. One essential tool was colabfold [23], which runs as a jupyter-notebook on the colab platform from google. Using MMseqs2 [24] for sequence searching makes building models extremely rapid, and the interface is straightforward for anyone to use. This shows the importance of open science.

The availability of the code enabled a large set of protein structures to be experimentally solved as the models were of sufficient quality to be used for molecular replacement [25]. One of the most impressive in this area is the model of the nucleopore complex [26].

AlphaFold2 was only developed to predict the structure of a single protein chain. However, it was soon realised that it was easy to trick the program into predicting the structure of dimers (or even higher multimers) by either assing a poly-Gly linker or simply changing the residue numbering. The first attempts had limited accuracy but once “paired” alignments linking the two (or more) chains were used, the performance increased significantly [27], and it can also be used to study [28] and design[29] protein-peptide interactions. This is important because protein-protein interactions are fundamental to understanding molecular functions. Further, this shows that AlphaFold2 has learned something about protein structure in general, i.e. it is not just a look-up table memorising all of PDB.

The success in predicting the structure of complexes led DeepMind to develop a version of AlphaFold aimed explicitly at this task. The first version (released in Nov 2021) could have been better (it tended to allow proteins to overlap). However, the second version (released in April 2022) [30] performs slightly better than the hacks using the original AlphaFold version [31]. It can accurately predict about half of all complexes up to 6 chains, and it is possible to predict the structure of larger complexes starting from predicted subcomplexes[32].

## AlphaFold2 clones/copies

In addition to the use of AlphaFold, the release of the code (and the detailed description) stimulated significant work to reproduce it. RoseTTAFold[22] was published at the same time as AlphaFold, but it is clear that the public description of AlphaFold strongly inspired their work. It contains some novelty, such as using a three-track network (one being SE(3) representation of the coordinates. However, initially, the performance was worse than AlphaFold. More importantly, it did not have learnt structural features as well as AlphaFold, as it could not be used for predicting protein-protein interactions. Later implementations of RoseTTAFold seem to rival AlphaFold in accuracy, and they have even enabled the prediction of proteins in complex with RNA or DNA[33]. RNA structure predictions using the ideas from AlphaFold is also possible[34]. Further, OpenFold[35] and UniFold[36] are clones of AlphaFold using another AI framework (PyTorch). They were released roughly one year after AlphaFold.

AlphaFold uses a multiple sequence alignment to predict the structure of a single protein. In some rare cases, the MSA is not needed, but generally, predictions without an MSA or very shallow MSAs are significantly worse. One way to improve the predictions for such proteins is to use a language model (i.e. a model trained to predict the sequence from the sequence). OmegaFold[37] omegafold and ESMfold[38] are two implementations that seem similar to AlphaFold but without using the MSA. However, whether the predictions using these models use a single sequence can be questioned. The performance is significantly worse for (orphan) proteins that do not have many homologs in the sequence databases, i.e. in some way, the language models appear to memorise the MSA. Esmfold is computationally efficient and has been used to predict the structure of all proteins from an extensive meta-genomics database[39].

# Use of AlphaFold for protein design.

In addition to providing the ability to predict the structure of proteins and complexes, AlphaFold has also been shown to be a valuable tool for protein design[29,40], and it has stimulated machine-learning tools for protein-ligand interactions[41]

# What did CASP15 teach us?

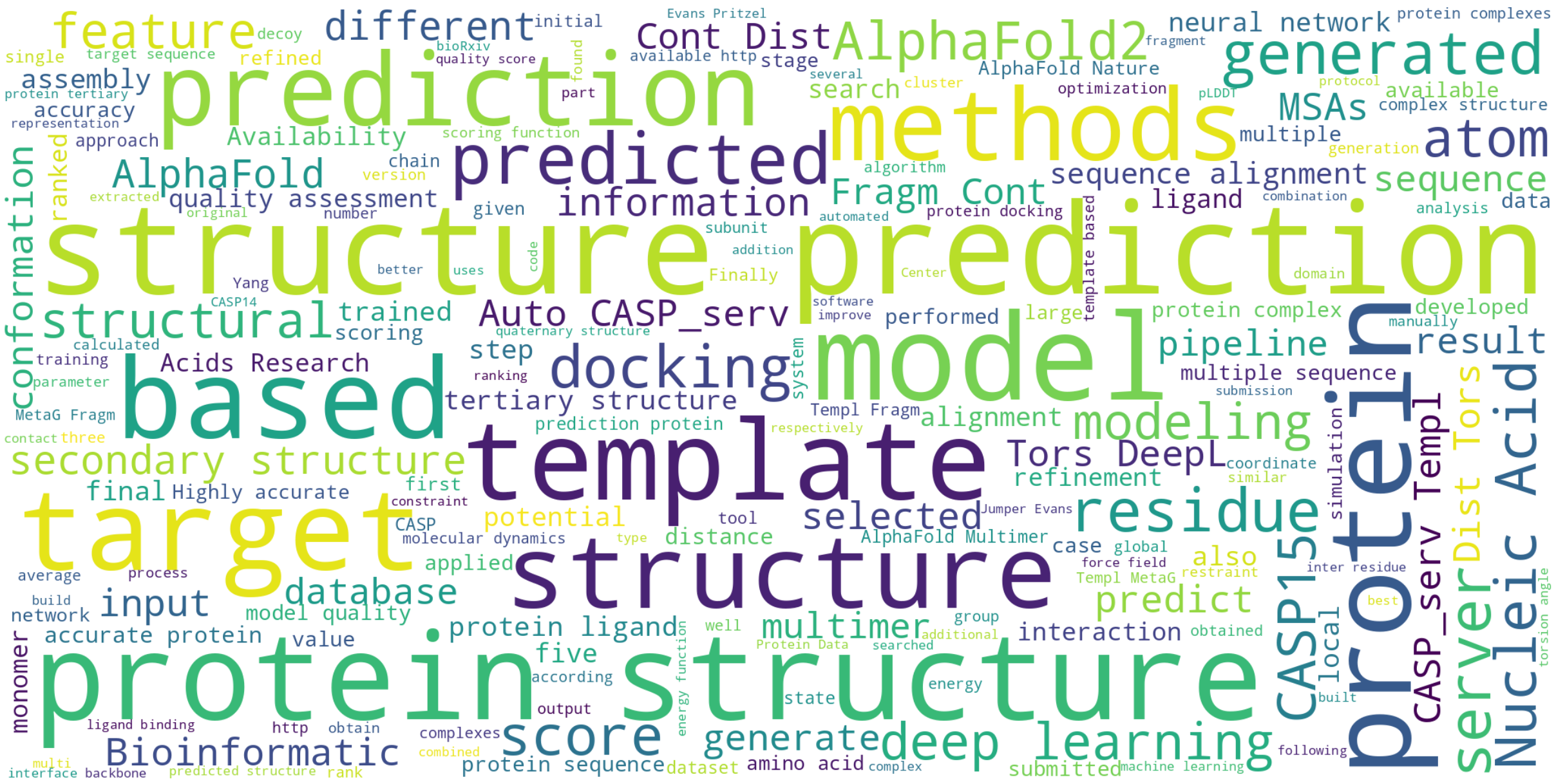

**Figure 1:** *Word cloud created from the abstract book of CASP15 shows that templates and AlphaFold are dominant factors when the participants describe their methods.*

Two years after the revolution provided by AlphaFold in CASP14, CASP15 was held. From reading the abstract, it can be seen that most groups used AlphaFold in one way or another (the term AlphaFold appears on average 1.4 times per page in the abstract book), Figure 1. Further, it was clear that the standard AlphaFold protocol performed better than more than half of the groups, Figure 2. Still, a few groups showed a substantial improvement over that method for individual proteins and protein assemblies. Five groups were selected to give presentations, and below I will explain how they could outperform AlphaFold. In short, there are two ways to improve over standard AlphaFold: to use templates more efficiently or to increase the sampling wither by using alternative methods to generate the MSAs or modify AlphaFold by using dropouts.

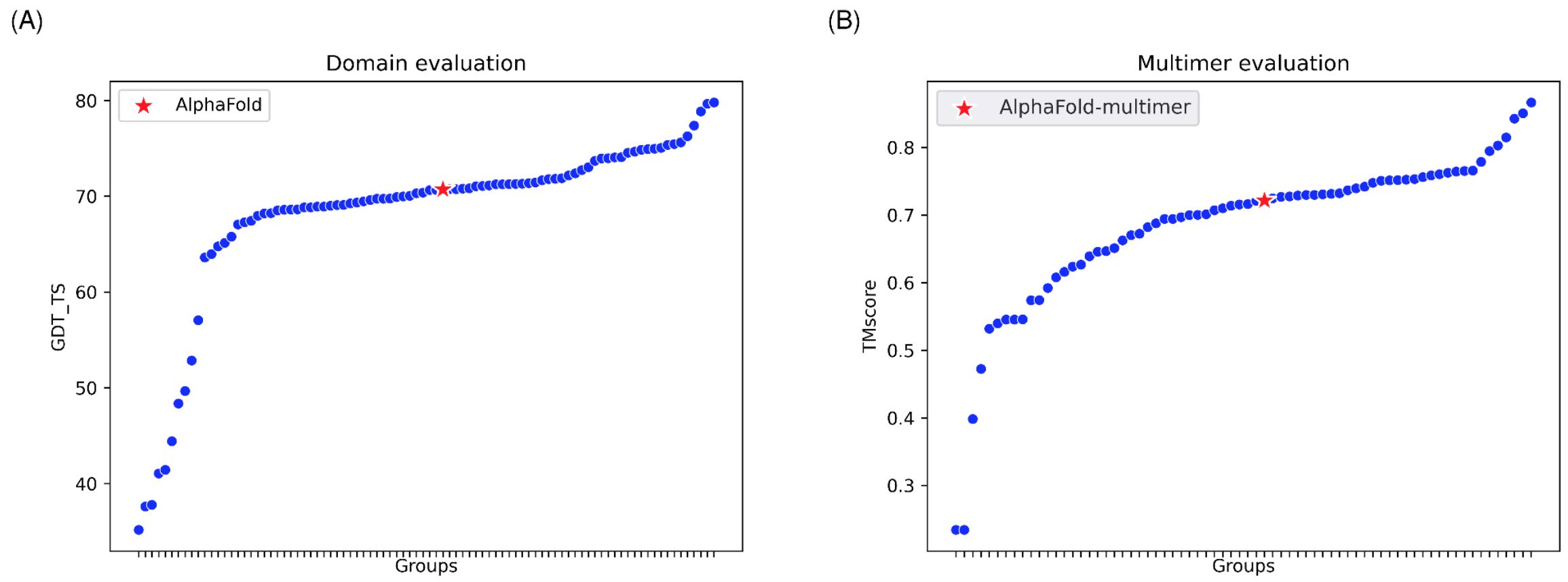

**Figure 2:** *Performance of default AlphaFold (red star) compared to all groups at CASP (blue dot). (A) Average GDT_TS for all monomeric targets, (B) average TMscore for all multimeric targets.*

Several groups produced improved multiple sequence alignments using DeepMSA2 [42] or by using more extensive databases. These MSAs are then used as input to AlphaFold or trRosetta, and the MSA that provides the best prediction (as measured by the estimated qualities) was then used for the predictions. This is an efficient way to generate an increased sampling directly from AlphaFold. An alternative way to generate structural diversity from AlphaFold is to use dropouts [43] or disabling pairing. Dropouts were used to produce thousands of models for each target.

Some groups used the predicted distances and other constraints from the AlphaFold models in other programs. Several groups used improved template searching by, e.g. using HHsearch [44] instead of HMMsearch [15,45]. When templates were identified, alternative modelling [15] or rigid body protein docking [46–48] protocols were used. Alternative scoring functions were used to identify the best models, including the one built into AlphaFold and scoring functions based on Voronoi surfaces [49] or Deep Learning.

# Conclusions

Since the release of AlphaFold in 2021, the field has shown steady progress, and hundreds of papers have already utilized AlphaFold. In CASP15, it is shown that running default AlphaFold on difficult targets, and the average TM-score is about 0.7. By using primarily increased sampling, it can be increased to about 0.8, i.e. for the vast majority of all proteins and protein complexes, AlphaFold can produce a model close to experimental quality.

# Funding

AE was funded by the Vetenskapsrådet Grant No. 2021-03979 and Knut and Alice Wallenberg foundation. The computations/data handling was enabled by the supercomputing resource Berzelius provided by National Supercomputer Centre at Linköping University and the Knut and Alice Wallenberg foundation and SNIC, grant No: SNIC 2021/5-297 and Berzelius-2021-29.